\documentclass[%
 reprint, 
 amsmath,amssymb,
 aps, physrev,
]{revtex4-2}

\usepackage{graphicx}
\usepackage{dcolumn}
\usepackage{bm}
\usepackage{hyperref}
\hypersetup{colorlinks=true, urlcolor=blue, linkcolor=blue, citecolor=blue}

\begin{document}

\title{\textbf{Sound Masking Strategies for Interference with Mosquito Hearing}}

\author{Justin Faber}
\email{faber@physics.ucla.edu}
\affiliation{Department of Physics \& Astronomy, University of California, Los Angeles, California, USA}

\author{Alexandros C Alampounti}
\affiliation{University College London Ear Institute: London, UK}

\author{Marcos Georgiades}
\affiliation{University College London Ear Institute: London, UK}

\author{Joerg T Albert}
\affiliation{University College London Ear Institute: London, UK}
\affiliation{Cluster of Excellence Hearing4all, Sensory Physiology \& Behaviour Group, Department for Neuroscience, School of Medicine and Health Sciences, Carl von Ossietzky Universität Oldenburg, Oldenburg, Germany}

\author{Dolores Bozovic}
\affiliation{Department of Physics \& Astronomy, University of California, Los Angeles, California, USA}
\affiliation{California NanoSystems Institute, University of California, Los Angeles, California, USA}

\date{\today}

\begin{abstract}
The use of auditory masking has long been of interest in psychoacoustics and for engineering purposes, in order to cover sounds that are disruptive to humans or to species whose habitats overlap with ours. In most cases, we seek to minimize the disturbances to the communication of wildlife. However, in the case of pathogen-carrying insects, we may want to maximize these disturbances as a way to control populations. In the current work, we explore candidate masking strategies for a generic model of active auditory systems and a model of the mosquito auditory system. For both models, we find that masks with all  acoustic power focused into just one or a few frequencies perform best. We propose that masks based on rapid frequency modulation are most effective for maximal disruption of information transfer and minimizing intelligibility. We hope that these results will serve to guide the avoidance or selection of possible acoustic signals for, respectively, maximizing or minimizing communication.

\end{abstract}

\maketitle

\section{Introduction}

When unwanted sound cannot be sufficiently attenuated or actively canceled, masking strategies are often used to obscure details and provide a more uniform acoustic environment \cite{kiddInformationalMasking2008, gustafssonMaskingSpeechAmplitudemodulated1994}. For applications aimed at minimizing distraction or improving sleep, using filtered noise as a sound mask is the typical strategy. Other applications include the masking of industrial noises in order to protect species susceptible to sound pollution \cite{clarkAcousticMaskingMarine2009, schmidtEcologyAcousticSignalling2015, derryberryPatternsSongNatural2016}. However, in some cases, we may want to maximize the disruption to animal behavior, particularly for species that are invasive or otherwise harmful.

Mosquitoes pose a significant threat to tropical and subtropical regions, resulting in over 700,000 annual human deaths and costing billions of dollars in healthcare expenses \cite{WHO2020}. Though advances have been made in developing pesticides and alternative strategies, the number of annual deaths continues to rise \cite{WHO2018}. Further, the rising global temperatures have led to the migration of the deadliest mosquito species to regions that were previously uninhabitable to them \cite{messinaManyProjectedFutures2015}. The need for additional control strategies is therefore evident. Some possible intervention strategies entail disrupting the mating or blood-feeding processes of these insects \cite{suAssessingAcousticBehaviour2020, andresBuzzkillTargetingMosquito2020}.

Mosquitoes mate at dusk, when males form swarms and seek out nearby females. Successful mating is dependent on the ability of a male to detect the sound of a female beating her wings. The male feather-like flagella are tuned approximately to the female's wingbeat frequency. Surprisingly, the active neuronal elements of the Johnston's organ are tuned to a much lower frequency \cite{booFineStructureScolopidia1975}. The nonlinear interaction between the female flight tone and the male's own wingbeats produce distortion products, which fall within the bandwidth of detection of the neuronal elements \cite{warrenSexRecognitionMidflight2009a, simoesRoleAcousticDistortion2016, simoesMaskingAuditoryBehaviour2018, suSexSpeciesSpecific2018}. This counterintuitive detection strategy has recently been modeled and proposed to be advantageous in a theoretical study \cite{faberMosquitoinspiredTheoreticalFramework2025a}. In the current work, we utilize theoretical models of mosquito hearing to find the optimal class of acoustic masks that would prevent a male from successfully detecting a female.

Measuring the performance of a biological detector relies on assumptions as to which components of the signal are meaningful. We must assume that the amplitude, frequency, phase, or some combination of these properties carries biologically relevant information. Alternatively, it could be modulations in these values that conveys information necessary for auditory detection. The goal of this study is not to determine the optimal sound mask for a specific species, but rather to find the optimal class of masks that would be generally effective in blocking active hearing. Therefore, we employ transfer entropy, an information theoretic measure that quantifies information imparted from one process to another \cite{Schreiber2000}. The transfer entropy does not rely on any assumptions of which components of the signal are meaningful. Rather, it reflects how much information about the stimulus is carried in the response of the receiver.

We note that sound detection and localization, along with sound scattering through the surrounding environment are all complex processes, which certainly contribute to shaping the optimal acoustic mask \cite{arthurMosquitoAedesAegypti2014, seoMechanismScalingWing2019, romerDirectionalHearingInsects2020, faberAntennalBasedStrategiesSound2025}. However, as these are specific to each environment and application of interest, we omit them from this study and focus on a single detector capturing a scalar signal. We do not discuss directional information that can be inferred from having multiple sensory elements, or by sensing the vector-valued velocity field of sound \cite{bennet-clarkAcousticsInsectSong1971, bennet-clarkSizeScaleEffects1998}. With this limitation established, we frame our problem as follows. Given a target acoustic signal of interest to an active auditory detector, what is the optimal acoustic mask that minimizes the information captured? Candidate masking signals can take any form we choose, with the constraint of fixed power input.

We use a Hopf oscillator as a generic model of auditory detection \cite{choeModelAmplificationHairbundle1998, eguiluzEssentialNonlinearitiesHearing2000, hudspethIntegratingActiveProcess2014, Reichenbach2014, alonsoAmplificationLocalCritical2025}, and a recently proposed model specific to the mosquito auditory system \cite{faberMosquitoinspiredTheoreticalFramework2025a}, based on a pair of Hopf oscillators and tuning to distortion products. We first present the performance of several mask categories: filtered noise, multi-tone signals, and stimuli with amplitude or frequency modulation. We then present a comparison of the best masks from each category. For both numerical models, we find filtered noise to be surprisingly ineffective, despite its common use in masking distraction-inducing sounds. For both models, we instead find that masks comprised of a single tone with constant amplitude and rapidly varying frequency perform best. Though these masks would likely be too unpleasant for human applications, they may provide acoustic-based strategies for interfering with the communication of pathogen carrying mosquitoes, as well as other unwanted species.

\section{Hopf oscillator (model 1)}

The active auditory system of vertebrates has long been described by a system poised near a critical point, on the verge of instability \cite{hudspethIntegratingActiveProcess2014}. In this regime, the system is highly sensitive to weak signals. This framework is elegantly captured by the normal form equation for the supercritical Hopf bifurcation \cite{choeModelAmplificationHairbundle1998, eguiluzEssentialNonlinearitiesHearing2000, alonsoAmplificationLocalCritical2025}. The state of this detector is described by a complex variable $z(t)$. The dynamics are governed by the first-order differential equation,

\begin{eqnarray}
\frac{dz}{dt} = (\mu + i\omega_0)z - |z|^2 z + F_f(t) + F_{mask}(t),
\end{eqnarray}

\noindent where $F_f(t)$ is the complex-valued signal of interest and $F_{mask}(t)$ represents the masking signal. In the absence of external stimulus ($F_f(t)=F_{mask}(t)=0$), this system exhibits autonomous limit-cycle oscillations for $\mu > 0$ and quiescent behavior for $\mu < 0$. The interface between these two regimes ($\mu=0$) defines the supercritical Hopf bifurcation. This system exhibits highest sensitivity near this bifurcation point. Autonomous oscillations occur at angular frequency, $\omega_0$, which coincides with the frequency of maximal sensitivity. Unless otherwise stated, we use $\omega_0 = 2\pi$ and $\mu = 0.1$, poising the system on the oscillatory side of the bifurcation.

\section{Hopf $\Rightarrow$ Hopf Mosquito Model (model 2)}

Insect hearing has also been shown to employ active processes \cite{gopfertActiveProcessesInsect2008, gopfertHearingInsects2016, nadrowskiAntennalHearingInsects2011, mhatreActiveAmplificationInsect2015, albertComparativeAspectsHearing2016}. We recently proposed a numerical model for the auditory system of the male mosquito \cite{faberMosquitoinspiredTheoreticalFramework2025a}. This model is comprised of two Hopf oscillators, one governing the mechanical tuning of the flagellum and the other governing the electrical tuning of the active neural elements. This model incorporates the evidence that these two tuning curves do not coincide. Rather, the electrical tuning aligns with a nonlinear distortion product produced by the simultaneous detection of the male and female wingbeats \cite{warrenSexRecognitionMidflight2009a, simoesRoleAcousticDistortion2016, simoesMaskingAuditoryBehaviour2018, suSexSpeciesSpecific2018}. The model takes the form

\begin{eqnarray*}
\frac{dz_1}{dt} = (\mu_1 + i\omega_1)z_1 - |z_1|^2 z_1 + F_f(t) + F_m(t) + F_{mask}(t)
\end{eqnarray*}

\begin{eqnarray}
\frac{dz_2}{dt} = (\mu_2 + i\omega_2)z_2 - |z_2|^2 z_2 + z_1(t),
\end{eqnarray}

\noindent where $z_1(t)$ and $z_2(t)$ represent the states of the mechanical and electrical components, respectively. $F_m(t)$ is the complex-valued stimulus from the male mosquito detecting his own wingbeats. Throughout this study, we keep both active elements in the oscillatory regime ($\mu_1 = \mu_2 = 0.1$). 

The acoustic signal of interest, $F_f(t)$ is confined to a narrow bandwidth, centered near the characteristic frequencies $\omega_0 = \omega_1 = 2\pi$. For the mosquito model, we must also include the male's own flight tone. Therefore, we use a pure tone of the form $F_m(t) = e^{i 3 \pi t}$, where we fix the amplitude at unity and use a frequency approximately 50\% higher than the female flight tones, consistent with experimental measurements \cite{suSexSpeciesSpecific2018}. The combination of male and female flight tones produces a nonlinear distortion product that falls near the characteristic frequency of the second oscillation ($\omega_2 = \pi$).

\section{Target Signal \& Transfer Entropy}

One limitation of our simplified framework, which focuses on a single detector with one spatial degree of freedom, is that it cannot extract directional information from the signals. This simplification, however, allows us to focus on just one task, namely, minimizing the detection of $F_f(t)$. Most common measures of detection sensitivity rely on assumptions of what properties of the signal carry biologically relevant information. Vector strength assumes that this information lies in the phase response of the detector. The linear response function assumes that the absolute amplitude of the response at the stimulus frequency carries meaningful information. Amplitude gain assumes the change in amplitude to be the key characteristic of a detector. To avoid these assumptions, we use the transfer entropy as our measure of detection.

\begin{figure}[t!]
\includegraphics[width=\columnwidth]{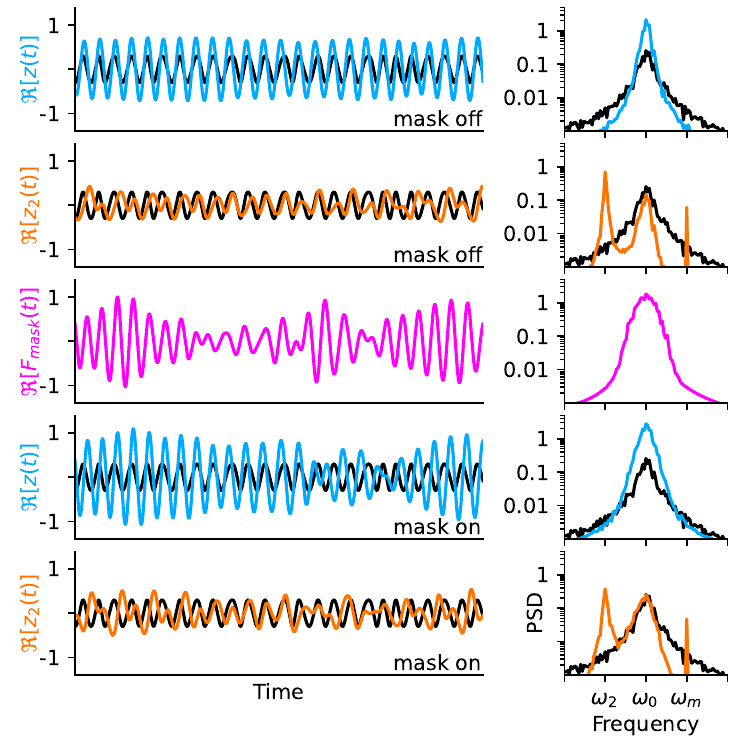}
\caption{Demonstration of the effects of a filtered-noise mask. The target signal and masking signal are shown in black and pink, respectively. The responses of model 1 (blue) and model 2 (orange) are shown in both the time and frequency domains. The top two rows show the responses in the absence of the masking signal, while the bottom two rows show the responses in the presence of the masking signal.}
\label{fig1}
\end{figure}

The transfer entropy quantifies how much information about the stimulus can be inferred from looking at the response of the detector (see Appendix A). In order to use this measure appropriately, we must use a target signal that constantly produces new information with time. We therefore use a sinusoid with fixed amplitude and stochastically modulated frequency, as this is the simplest nonstationary signal that approximates the flight tones of insects. Our target signal takes the form,

\begin{eqnarray} \label{eq:Ff1}
F_f(t) = f_0 e^{i\phi(t)}
\end{eqnarray}

\begin{eqnarray}  \label{eq:Ff2}
\phi(t) = \int_{-\infty}^t \Big(\omega_0 + \eta(t)\Big) dt,
\end{eqnarray}

\noindent where $\omega_0 = \omega_1 = 2\pi$ is the mean instantaneous frequency and $\eta(t)$ is a stochastic variable with standard deviation $0.2 \times \omega_0$. We consider slow, smooth modulations in this variable by using pink-noise statistics. The power spectrum of $\eta(t)$ is nonzero and uniform from $0$ to $\omega_{mask}$ and zero elsewhere. In FIG. \ref{fig1}, we show an example of a candidate masking signal and how it influences the response of each of the two models to the target signal.

To interfere with detection, we develop a masking strategy based on our knowledge of the statistics of the target signal. We know the mean and standard deviation of $\eta(t)$, but not the values at each point in time. If we knew the full time series, we could construct a masking signal that perfectly cancels the target signal. Likewise, for blocking communication between mosquitoes, we know only the statistics of the tones at which they communicate.

With our target signal defined, and our goal focused on minimizing transfer entropy from $F_f(t)$ to $z(t)$ (model 1) or $F_f(t)$ to $z_2(t)$ (model 2), we now define our normalization constant for the masking signal. Detection can be masked trivially by increasing the amplitude of the mask arbitrarily high. To avoid trivial solutions and to find a mask that can be effective for the greatest distance from the acoustic masking source, we impose the constraint of fixed power input,

\begin{eqnarray}
P_m = \int_{-\infty}^{\infty} F_{mask}^*(\tau) F_{mask}(\tau) d\tau,
\end{eqnarray}

\noindent where $*$ denotes the complex conjugate. We also define the mask-to-signal ratio to be the power ratio of the mask to the target signal, $P_m/f_0^2$. Unless otherwise stated, we use $f_0=0.3$ and a mask-to-signal ratio of $1$. We now explore several categories of possible masking strategies.

\section{Filtered-Noise Masks}

\begin{figure}[t!]
\includegraphics[width=\columnwidth]{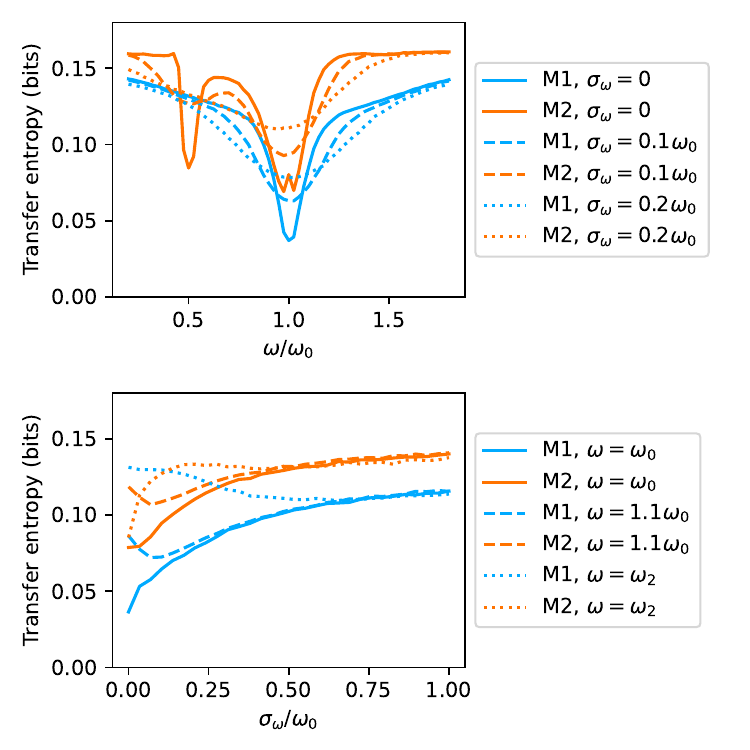}
\caption{\textbf{Filtered-noise mask.} The effects of the masking parameters, $\omega$ and $\sigma_{\omega}$, on the transfer entropy from $F_f(t)$ to $z(t)$ (model 1: M1), and from $F_f(t)$ to $z_2(t)$ (model 2: M2).}
\label{fig2}
\end{figure}

A natural starting point for designing an acoustic mask is to attempt to corrupt communication using stochastic white noise. However, since our target signal is confined to a bandwidth surrounding $\omega_0$, we may want to filter this noise, so as to focus more acoustic power onto the bandwidth of communication. To produce a filtered-noise mask, we first generate Gaussian white noise. We then apply a bandpass, Gaussian-window filter in the frequency domain, where $\omega$ and $\sigma_{\omega}$ are the mean and standard deviation of the window function. Finally, we transform back to the time domain and rescale the signal variance to be $P_m$. In FIG. \ref{fig2}, we show how the transfer entropy for both models depends on $\omega$ and $\sigma_{\omega}$. Note that in the limit of $\sigma_{\omega} \rightarrow \infty$, the mask is Gaussian white noise. However, in the limit of $\sigma_{\omega} \rightarrow 0$, the mask is a pure tone.

Surprisingly, we find that the transfer entropy is not minimized when the bandwidth of the mask matches the bandwidth of the target signal. Instead, for both models, minimal transfer entropy is found when $\sigma_{\omega} = 0$ and $\omega \approx \omega_0$, indicating that a pure tone near the center frequency is more effective at corrupting communication than stochastic noise. For model 2, there is an additional minimum at $\omega \approx \omega_2$, indicating that an effective masking strategy could entail corrupting communication at either characteristic frequency of the two oscillators. For mosquitoes, this regime corresponds to focusing the acoustic power near the characteristic frequency of either the flagellum or the neuronal elements.

\section{Two-tone Masks}

\begin{figure}[t!]
\includegraphics[width=\columnwidth]{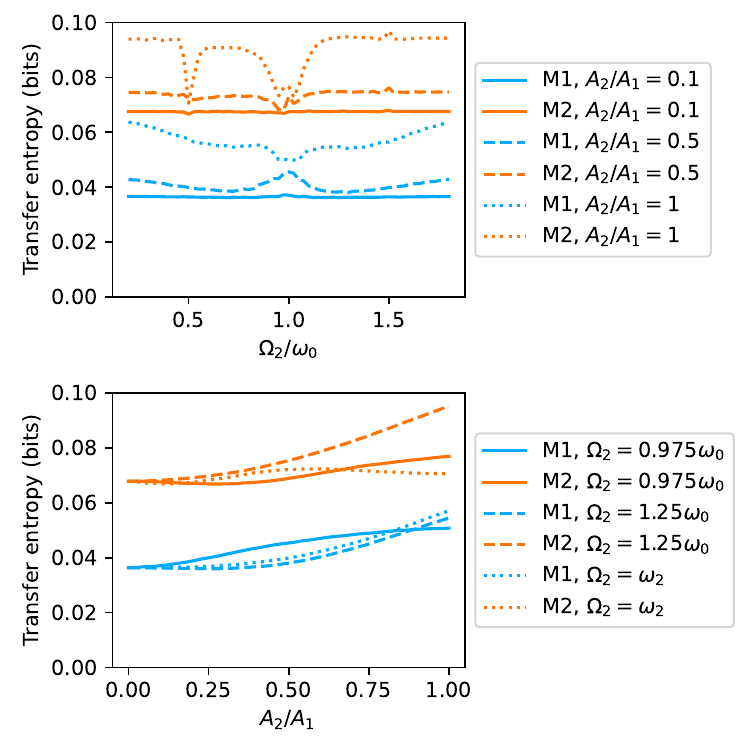}
\caption{\textbf{Two-tone mask.} The effects of the masking parameters, $\Omega_2$ and $A_2/A_1$, on the transfer entropy from $F_f(t)$ to $z(t)$ (model 1: M1), and from $F_f(t)$ to $z_2(t)$ (model 2: M2).}
\label{fig3}
\end{figure}

We now consider acoustic masks comprised of just two pure tones, rather than a dense bandwidth of frequencies. Since a pure tone mask was most effective in the previous section, a natural extension is to test if there is an advantage to splitting the power into two frequency components. We consider masks with the functional form,

\begin{eqnarray}
F_{mask}(t) = A_1 e^{i\Omega_1 t} + A_2 e^{i\Omega_2 t},
\end{eqnarray}

\noindent where $A_1$ and $A_2$ represent the amplitudes, and $\Omega_1$ and $\Omega_2$ the frequencies of the two tones. Without loss of generality, we let $A_1 \geq A_2$. We fix $\Omega_1 = 1.01\times\omega_0$, setting the stronger tone near the characteristic frequency. We then vary the ratio, $A_2/A_1$, and $\Omega_2$ and find the effects on the transfer entropy. In FIG. \ref{fig3}, we observe that, for both models, there are shallow minima in the plots of transfer entropy as a function of $A_2/A_1$, with hardly any advantage gained by adding a second tone.

\section{Amplitude-Modulation (AM) Masks}

\begin{figure}[t!]
\includegraphics[width=\columnwidth]{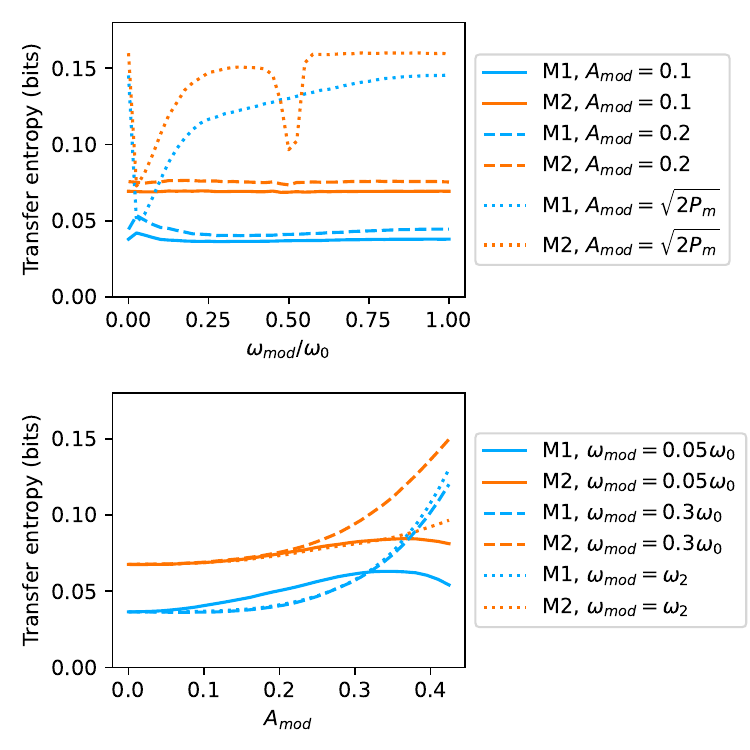}
\caption{\textbf{Amplitude-modulation (AM) mask.} The effects of the masking parameters, $\omega_{mod}$ and $A_{mod}$, on the transfer entropy from $F_f(t)$ to $z(t)$ (model 1: M1), and from $F_f(t)$ to $z_2(t)$ (model 2: M2).}
\label{fig4}
\end{figure}

We now consider AM sound masks, where we sinusoidally modulate the amplitude of a stimulus tone near the characteristic frequency. This mask takes the form,

\begin{eqnarray}
F_{mask}(t) = \big(A_0 + A_{mod}\sin(\omega_{mod}t)\big) e^{i\Omega t},
\end{eqnarray}

\noindent where $\Omega = 1.01\times\omega_0$. This mask can also be regarded as a 3-tone stimulus. Using trigonometric identities, we can express $F_{mask}(t)$ as a sum of three pure tones at frequencies $\Omega$, $\Omega - \omega_{mod}$, and $\Omega + \omega_{mod}$. We vary $\omega_{mod}$ and $A_{mod}$ and find the effects on the transfer entropy. Note that our constraint of fixed mask power uniquely determines $A_0$ for every choice of $A_{mod}$. We find that amplitude modulations do not noticeably improve the effectiveness of the mask (FIG. \ref{fig4}). For model 1, there is once again a very shallow minimum near $A_{mod}=0.1$. However, the minimal transfer entropy for model 2 is found at $A_{mod}=0$.

\section{Frequency-Modulation (FM) Masks}

We now explore masks based on frequency modulation. We consider the case where all of the acoustic power is focused into one frequency, but that frequency is modulated with time, so as to corrupt a wider bandwidth. We use a sound mask of the form,

\begin{eqnarray} \label{eq:fm1}
F_f(t) = \sqrt{P_m} e^{i\psi(t)}
\end{eqnarray}

\begin{eqnarray} \label{eq:fm2}
\psi(t) = \int_{-\infty}^t \Big(\Omega + m(t)\Big) dt,
\end{eqnarray}

\noindent where $\Omega = \omega_0 = \omega_1$ is the carrier frequency and $m(t)$ is the modulator. We consider periodic modulators that follow a power-law increase,

\begin{eqnarray} \label{eq:fm4}
m(t) = A_{mod} (2f_{mod}t - 1) |2f_{mod}t - 1|^{\alpha-1},
\end{eqnarray}

\noindent over each modulation period, $t=\frac{n}{f_{mod}}$ to $t=\frac{n+1}{f_{mod}}$ for any $n$. Note that $m(t)$ spans $-A_{mod}$ to $+A_{mod}$ over each modulation period. 

$\alpha$ is a real-valued, non-negative free parameter characterizing the power-law growth. For large values of $\alpha$, the frequency sweep spends more time near the center frequency, $\omega_0$, while for small values, the sweep spends more time time near the end points of the sweep, $\omega_0 \pm A_{mod}$. For $\alpha = 1$, the frequency increases linearly from $\omega_0-A_{mod}$ to $\omega_0+A_{mod}$ and then repeats. This is known as sawtooth frequency modulation. For $\alpha = 0$, $m(t)$ becomes a square wave, which abruptly switches between the two frequency extrema. We vary $\alpha$, $A_{mod}$, and $f_{mod}$ so as to test a wide variety of frequency modulators.

\begin{figure}[t!]
\includegraphics[width=\columnwidth]{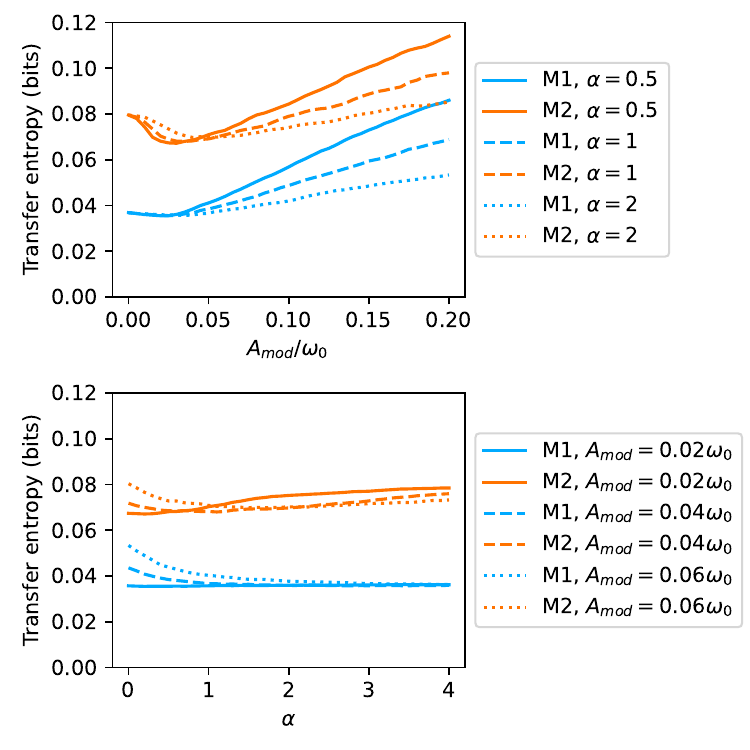}
\caption{\textbf{Power-law frequency-modulation (FM) mask.} The effects of the masking parameters, $\omega_{mod}$ and $A_{mod}$, on the transfer entropy from $F_f(t)$ to $z(t)$ (model 1: M1), and from $F_f(t)$ to $z_2(t)$ (model 2: M2).}
\label{fig5}
\end{figure}

In FIG. \ref{fig5}, we show the effects of varying $A_{mod}$ and $\alpha$. The transfer entropy is minimal with small modulations in the stimulus frequency. Further, the results are rather insensitive to the value of $\alpha$. In Appendix B, we show the results from FM masks with sinusoidal, sawtooth, and square-wave modulators. We further show the effects of varying $f_{mod}$.

\section{Comparison of all mask types}

\begin{figure}[t!]
\includegraphics[width=\columnwidth]{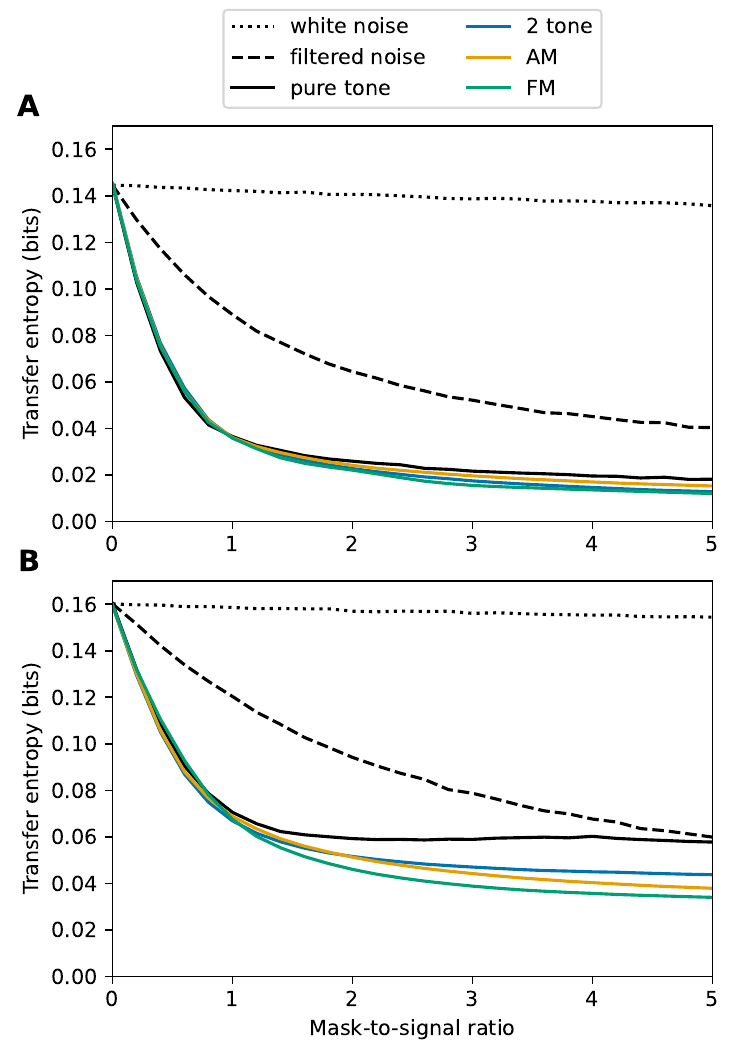}
\caption{\textbf{Comparison across mask strength.} Transfer entropy dependence on masking strength for model 1 (A) and model 2 (B). Mask parameters in (A) are $\omega=\omega_0$ and $\sigma_{\omega}=\infty$ (for white noise), $\omega=\omega_0$ and $\sigma_{\omega}=0.3\omega_0$ (for filtered noise), $\omega=\omega_0$ and $\sigma_{\omega}=0$ (for pure tone), $\Omega_1 = 1.01\omega_0$, $\Omega_2 = 1.25\omega_0$, and $A_2/A_1 = 0.25$ (for 2 tone), $\omega_{mod} = 0.3\omega_0$ and $A_{mod}=0.07$ (for AM), $\omega_{mod} = 0.22\omega_0$ and $A_{mod} = 0.08\omega_0$ (for sinusoidal FM). Mask parameters in (B) are $\omega=\omega_0$ and $\sigma_{\omega}=\infty$ (for white noise), $\omega=\omega_0$ and $\sigma_{\omega}=0.3\omega_0$ (for filtered noise), $\omega = 0.975\omega_0$ and $\sigma_{\omega}=0$ (for pure tone), $\Omega_1 = 1.01\omega_0$, $\Omega_2 = 0.975\omega_0$, and $A_2/A_1 = 0.25$ (for 2 tone), $\omega_{mod} = 0.5\omega_0$ and $A_{mod}=0.1$ (for AM), $\omega_{mod} = 0.5\omega_0$ and $A_{mod} = 0.25\omega_0$ (for sinusoidal FM).}
\label{fig6}
\end{figure}

Having explored a variety of sound masks and identified the optimal parameter ranges, we now compare the best masks from each of the categories. We note that all tests so far were performed with a mask power equal to that of the target signal. Therefore, our mask-to-signal ratio has so far been $P_m/f_0^2 = 1$. However, this ratio will differ from $1$, depending on the application, the distance from the sound sources, and the acoustic power available. Therefore, we vary the signal-to-mask ratio and assess the performance of the optimal mask from each of the four categories (FIG. \ref{fig6}). For comparison, we also show the performance of masks comprised of white noise and bandpass filtered noise.

We find that white noise and filtered noise perform poorly as sound masks in terms of transfer entropy reduction. However, the most effective masks are those with all of the acoustic power focused into just one or a few frequencies at any given time. We find FM masks to be the most effective, though the pure tone, two tone, and AM masks are comparable. It is also worth noting that for all mask types and for all mask-to-signal ratios, the transfer entropy to model 2 is greater than that of model 1. This highlights the high sensitivity and robustness of the detection scheme employed by male mosquitoes \cite{faberMosquitoinspiredTheoreticalFramework2025a}. Despite the loss in signal power associated with using distortion products instead of primary tones, the Hopf $\Rightarrow$ Hopf detector captures more information from external signals than does the single Hopf oscillator, even in the presence of sound masking.

\section{Discussion}

Previous studies have attempted to reduce mosquito populations using acoustic traps \cite{andresBuzzkillTargetingMosquito2020}. The strategy behind this approach is to lure male mosquitoes by presenting artificial female flight tones. These methods have shown promising results in both laboratory \cite{jakheteWingbeatFrequencySweepVisual2017} and field \cite{ogawaFieldStudyAcoustic1988, ikeshojiImpactInsecticidetreatedSound1990, stoneDeterminantsMaleAedes2013, johnsonSirensSongExploitation2016, rohdeWaterproofLowcostLongbatterylife2019} settings, with improved success when combined with carbon dioxide release and using live animals as bait. There have also been attempts to control the behavior of mosquitoes using disruptive sounds, with studies showing that mating and blood-feeding activity are both reduced in the presence of loud music \cite{diengElectronicSongScary2019, niImpactDiverseMusical2025}. However, the practicality of these strategies for widespread use remains unclear, as does the nature of the most effective sound for disrupting mosquito hearing.

In the current study, we aimed to answer the question of which sound mask is theoretically most effective. We used a Hopf oscillator as a generic model of active additory systems. We also used a recently proposed numerical model of the mosquito auditory system. We compared several classes of masking strategies on the two models, showing that rapid frequency sweeps are most effective in corrupting signal detection. Further, we varied the masking power, so as to explore the robustness of these results for various application. We found that, in all regimes, it is favorable to have all of the acoustic energy focused into just one or a few frequencies at any given time. Further improvement can be seen with a single tone of rapidly varying frequency.

This method is known in radar jamming science as \textit{barrage jamming} \cite{SkolnikRadar1970}. This is an effective strategy that allows the jammer to corrupt a wide bandwidth by rapidly modulating one stimulus frequency. Consistent with these prior applications, we found it to be the most effective mask, for both models of the auditory system. Interestingly, the song used in a previous study for disrupting mosquito behavior contains frequency sweeps \cite{diengElectronicSongScary2019}. Another advantage of the frequency sweep is that the statistics of target signal do not need to be precisely known, as long as its bandwidth overlaps with the range of the frequency sweep. Mosquito flight tones and the frequency of highest sensitivity vary between species \cite{suSexSpeciesSpecific2018}, and even vary with time, following a circadian cycle \cite{somersHittingRightNote2022}. Hence, this insensitivity to precise masking parameters could be useful for designing acoustic masks. We therefore propose that the use of rapid acoustic frequency sweeps could provide a practical solution for controlling the mating and blood-feeding behavior of mosquitoes.

We note that the masks presented in this study represent only a small subset of all functional forms possible. Because bandwidth noise masks proved ineffective, and the addition of a second or third pure tone did not show much improvement over a single tone, we were led to rule out masks that do not have all power focused into a single frequency for any short interval of time. This left us with just one class of mask, namely, FM masks described by Eqs. \ref{eq:fm1} and \ref{eq:fm2}. However, we note that more effective masks may exist. Future work entails exploring masks that have both a modulated amplitude and frequency. Other masks that could be explored entail FM mask in which the modulator abruptly jumps between many frequencies, possibly in a stochastic manner. Future work also entails using machine learning algorithms to speed up the exploration of the vast parameter space of masking signals, with the challenge of parameterizing a broadest range of signal classes while using the fewest parameters possible. While future improvements may further enhance the ability to corrupt auditory detection by mosquitoes, the current work demonstrates that signals with rapid frequency modulation may provide a plausible approach for population control of harmful species.

\section*{Acknowledgments}
This work was supported by a grant from the Biotechnology and Biological Sciences Research Council, UK (BBSRC, BB/V007866/1 to J.T.A.) and a grant from The Human Frontier Science Program (HFSP grant RGP0033/2021 to J.T.A. and D.B.).

\section*{Data Availability}
The Python code for performing the analysis and generating the figures is publicly available online: \cite{DataAvailability}.

\bibliography{Bibliography}

\providecommand{\noopsort}[1]{}\providecommand{\singleletter}[1]{#1}%
\begin{thebibliography}{45}%
\makeatletter
\providecommand \@ifxundefined [1]{%
 \@ifx{#1\undefined}
}%
\providecommand \@ifnum [1]{%
 \ifnum #1\expandafter \@firstoftwo
 \else \expandafter \@secondoftwo
 \fi
}%
\providecommand \@ifx [1]{%
 \ifx #1\expandafter \@firstoftwo
 \else \expandafter \@secondoftwo
 \fi
}%
\providecommand \natexlab [1]{#1}%
\providecommand \enquote  [1]{``#1''}%
\providecommand \bibnamefont  [1]{#1}%
\providecommand \bibfnamefont [1]{#1}%
\providecommand \citenamefont [1]{#1}%
\providecommand \href@noop [0]{\@secondoftwo}%
\providecommand \href [0]{\begingroup \@sanitize@url \@href}%
\providecommand \@href[1]{\@@startlink{#1}\@@href}%
\providecommand \@@href[1]{\endgroup#1\@@endlink}%
\providecommand \@sanitize@url [0]{\catcode `\\12\catcode `\$12\catcode
  `\&12\catcode `\#12\catcode `\^12\catcode `\_12\catcode `\%12\relax}%
\providecommand \@@startlink[1]{}%
\providecommand \@@endlink[0]{}%
\providecommand \url  [0]{\begingroup\@sanitize@url \@url }%
\providecommand \@url [1]{\endgroup\@href {#1}{\urlprefix }}%
\providecommand \urlprefix  [0]{URL }%
\providecommand \Eprint [0]{\href }%
\providecommand \doibase [0]{https://doi.org/}%
\providecommand \selectlanguage [0]{\@gobble}%
\providecommand \bibinfo  [0]{\@secondoftwo}%
\providecommand \bibfield  [0]{\@secondoftwo}%
\providecommand \translation [1]{[#1]}%
\providecommand \BibitemOpen [0]{}%
\providecommand \bibitemStop [0]{}%
\providecommand \bibitemNoStop [0]{.\EOS\space}%
\providecommand \EOS [0]{\spacefactor3000\relax}%
\providecommand \BibitemShut  [1]{\csname bibitem#1\endcsname}%
\let\auto@bib@innerbib\@empty
\bibitem [{\citenamefont {Kidd}\ \emph {et~al.}(2008)\citenamefont {Kidd},
  \citenamefont {Mason}, \citenamefont {Richards}, \citenamefont {Gallun},\
  and\ \citenamefont {Durlach}}]{kiddInformationalMasking2008}%
  \BibitemOpen
  \bibfield  {author} {\bibinfo {author} {\bibfnamefont {G.}~\bibnamefont
  {Kidd}}, \bibinfo {author} {\bibfnamefont {C.~R.}\ \bibnamefont {Mason}},
  \bibinfo {author} {\bibfnamefont {V.~M.}\ \bibnamefont {Richards}}, \bibinfo
  {author} {\bibfnamefont {F.~J.}\ \bibnamefont {Gallun}},\ and\ \bibinfo
  {author} {\bibfnamefont {N.~I.}\ \bibnamefont {Durlach}},\ }\bibfield
  {title} {\bibinfo {title} {Informational {{Masking}}},\ }in\ \href
  {https://doi.org/10.1007/978-0-387-71305-2_6} {\emph {\bibinfo {booktitle}
  {Auditory {{Perception}} of {{Sound Sources}}}}},\ \bibinfo {editor} {edited
  by\ \bibinfo {editor} {\bibfnamefont {W.~A.}\ \bibnamefont {Yost}}, \bibinfo
  {editor} {\bibfnamefont {A.~N.}\ \bibnamefont {Popper}},\ and\ \bibinfo
  {editor} {\bibfnamefont {R.~R.}\ \bibnamefont {Fay}}}\ (\bibinfo  {publisher}
  {Springer US},\ \bibinfo {year} {2008})\ pp.\ \bibinfo {pages}
  {143--189}\BibitemShut {NoStop}%
\bibitem [{\citenamefont {Gustafsson}\ and\ \citenamefont
  {Arlinger}(1994)}]{gustafssonMaskingSpeechAmplitudemodulated1994}%
  \BibitemOpen
  \bibfield  {author} {\bibinfo {author} {\bibfnamefont {H.~{\AA}.}\
  \bibnamefont {Gustafsson}}\ and\ \bibinfo {author} {\bibfnamefont {S.~D.}\
  \bibnamefont {Arlinger}},\ }\bibfield  {title} {\bibinfo {title} {Masking of
  speech by amplitude-modulated noise},\ }\href
  {https://doi.org/10.1121/1.408346} {\bibfield  {journal} {\bibinfo  {journal}
  {JASA}\ }\textbf {\bibinfo {volume} {95}},\ \bibinfo {pages} {518} (\bibinfo
  {year} {1994})}\BibitemShut {NoStop}%
\bibitem [{\citenamefont {Clark}\ \emph {et~al.}(2009)\citenamefont {Clark},
  \citenamefont {Ellison}, \citenamefont {Southall}, \citenamefont {Hatch},
  \citenamefont {Parijs}, \citenamefont {Frankel},\ and\ \citenamefont
  {Ponirakis}}]{clarkAcousticMaskingMarine2009}%
  \BibitemOpen
  \bibfield  {author} {\bibinfo {author} {\bibfnamefont {C.~W.}\ \bibnamefont
  {Clark}}, \bibinfo {author} {\bibfnamefont {W.~T.}\ \bibnamefont {Ellison}},
  \bibinfo {author} {\bibfnamefont {B.~L.}\ \bibnamefont {Southall}}, \bibinfo
  {author} {\bibfnamefont {L.}~\bibnamefont {Hatch}}, \bibinfo {author}
  {\bibfnamefont {S.~M.~V.}\ \bibnamefont {Parijs}}, \bibinfo {author}
  {\bibfnamefont {A.}~\bibnamefont {Frankel}},\ and\ \bibinfo {author}
  {\bibfnamefont {D.}~\bibnamefont {Ponirakis}},\ }\bibfield  {title} {\bibinfo
  {title} {Acoustic masking in marine ecosystems: Intuitions, analysis, and
  implication},\ }\href {https://doi.org/10.3354/meps08402} {\bibfield
  {journal} {\bibinfo  {journal} {Mar. Ecol. Prog. Ser.}\ }\textbf {\bibinfo
  {volume} {395}},\ \bibinfo {pages} {201} (\bibinfo {year}
  {2009})}\BibitemShut {NoStop}%
\bibitem [{\citenamefont {Schmidt}\ and\ \citenamefont
  {Balakrishnan}(2015)}]{schmidtEcologyAcousticSignalling2015}%
  \BibitemOpen
  \bibfield  {author} {\bibinfo {author} {\bibfnamefont {A.~K.~D.}\
  \bibnamefont {Schmidt}}\ and\ \bibinfo {author} {\bibfnamefont
  {R.}~\bibnamefont {Balakrishnan}},\ }\bibfield  {title} {\bibinfo {title}
  {Ecology of acoustic signalling and the problem of masking interference in
  insects},\ }\href {https://doi.org/10.1007/s00359-014-0955-6} {\bibfield
  {journal} {\bibinfo  {journal} {J. Comp. Physiol. A}\ }\textbf {\bibinfo
  {volume} {201}},\ \bibinfo {pages} {133} (\bibinfo {year}
  {2015})}\BibitemShut {NoStop}%
\bibitem [{\citenamefont {Derryberry}\ \emph {et~al.}(2016)\citenamefont
  {Derryberry}, \citenamefont {Danner}, \citenamefont {Danner}, \citenamefont
  {Derryberry}, \citenamefont {Phillips}, \citenamefont {Lipshutz},
  \citenamefont {Gentry},\ and\ \citenamefont
  {Luther}}]{derryberryPatternsSongNatural2016}%
  \BibitemOpen
  \bibfield  {author} {\bibinfo {author} {\bibfnamefont {E.~P.}\ \bibnamefont
  {Derryberry}}, \bibinfo {author} {\bibfnamefont {R.~M.}\ \bibnamefont
  {Danner}}, \bibinfo {author} {\bibfnamefont {J.~E.}\ \bibnamefont {Danner}},
  \bibinfo {author} {\bibfnamefont {G.~E.}\ \bibnamefont {Derryberry}},
  \bibinfo {author} {\bibfnamefont {J.~N.}\ \bibnamefont {Phillips}}, \bibinfo
  {author} {\bibfnamefont {S.~E.}\ \bibnamefont {Lipshutz}}, \bibinfo {author}
  {\bibfnamefont {K.}~\bibnamefont {Gentry}},\ and\ \bibinfo {author}
  {\bibfnamefont {D.~A.}\ \bibnamefont {Luther}},\ }\bibfield  {title}
  {\bibinfo {title} {Patterns of {{Song}} across {{Natural}} and
  {{Anthropogenic Soundscapes Suggest That White-Crowned Sparrows Minimize
  Acoustic Masking}} and {{Maximize Signal Content}}},\ }\href
  {https://doi.org/10.1371/journal.pone.0154456} {\bibfield  {journal}
  {\bibinfo  {journal} {PLOS ONE}\ }\textbf {\bibinfo {volume} {11}},\ \bibinfo
  {pages} {e0154456} (\bibinfo {year} {2016})}\BibitemShut {NoStop}%
\bibitem [{\citenamefont {Organization}(2020)}]{WHO2020}%
  \BibitemOpen
  \bibfield  {author} {\bibinfo {author} {\bibfnamefont {W.~H.}\ \bibnamefont
  {Organization}},\ }\href
  {https://www.who.int/news-room/fact-sheets/detail/vector-borne-diseases}
  {\bibinfo {title} {Vector-borne diseases}} (\bibinfo {year}
  {2020})\BibitemShut {NoStop}%
\bibitem [{\citenamefont {Organization}(2018)}]{WHO2018}%
  \BibitemOpen
  \bibfield  {author} {\bibinfo {author} {\bibfnamefont {W.~H.}\ \bibnamefont
  {Organization}},\ }\href
  {https://iris.who.int/bitstream/handle/10665/272533/9789241514057-eng.pdfs}
  {\bibinfo {title} {Global report on insecticide resistance in malaria
  vectors: 2010–2016}} (\bibinfo {year} {2018})\BibitemShut {NoStop}%
\bibitem [{\citenamefont {Messina}\ \emph {et~al.}(2015)\citenamefont
  {Messina}, \citenamefont {Brady}, \citenamefont {Pigott}, \citenamefont
  {Golding}, \citenamefont {Kraemer}, \citenamefont {Scott}, \citenamefont
  {Wint}, \citenamefont {Smith},\ and\ \citenamefont
  {Hay}}]{messinaManyProjectedFutures2015}%
  \BibitemOpen
  \bibfield  {author} {\bibinfo {author} {\bibfnamefont {J.~P.}\ \bibnamefont
  {Messina}}, \bibinfo {author} {\bibfnamefont {O.~J.}\ \bibnamefont {Brady}},
  \bibinfo {author} {\bibfnamefont {D.~M.}\ \bibnamefont {Pigott}}, \bibinfo
  {author} {\bibfnamefont {N.}~\bibnamefont {Golding}}, \bibinfo {author}
  {\bibfnamefont {M.~U.~G.}\ \bibnamefont {Kraemer}}, \bibinfo {author}
  {\bibfnamefont {T.~W.}\ \bibnamefont {Scott}}, \bibinfo {author}
  {\bibfnamefont {G.~R.~W.}\ \bibnamefont {Wint}}, \bibinfo {author}
  {\bibfnamefont {D.~L.}\ \bibnamefont {Smith}},\ and\ \bibinfo {author}
  {\bibfnamefont {S.~I.}\ \bibnamefont {Hay}},\ }\bibfield  {title} {\bibinfo
  {title} {The many projected futures of dengue},\ }\href
  {https://doi.org/10.1038/nrmicro3430} {\bibfield  {journal} {\bibinfo
  {journal} {Nat. Rev. Microbiol.}\ }\textbf {\bibinfo {volume} {13}},\
  \bibinfo {pages} {230} (\bibinfo {year} {2015})}\BibitemShut {NoStop}%
\bibitem [{\citenamefont {Su}\ \emph {et~al.}(2020)\citenamefont {Su},
  \citenamefont {Georgiades}, \citenamefont {Bagi}, \citenamefont {Kyrou},
  \citenamefont {Crisanti},\ and\ \citenamefont
  {Albert}}]{suAssessingAcousticBehaviour2020}%
  \BibitemOpen
  \bibfield  {author} {\bibinfo {author} {\bibfnamefont {M.~P.}\ \bibnamefont
  {Su}}, \bibinfo {author} {\bibfnamefont {M.}~\bibnamefont {Georgiades}},
  \bibinfo {author} {\bibfnamefont {J.}~\bibnamefont {Bagi}}, \bibinfo {author}
  {\bibfnamefont {K.}~\bibnamefont {Kyrou}}, \bibinfo {author} {\bibfnamefont
  {A.}~\bibnamefont {Crisanti}},\ and\ \bibinfo {author} {\bibfnamefont
  {J.~T.}\ \bibnamefont {Albert}},\ }\bibfield  {title} {\bibinfo {title}
  {Assessing the acoustic behaviour of {{Anopheles}} gambiae (s.l.) {{dsxF}}
  mutants: implications for vector control},\ }\href
  {https://doi.org/10.1186/s13071-020-04382-x} {\bibfield  {journal} {\bibinfo
  {journal} {Parasit. Vectors}\ }\textbf {\bibinfo {volume} {13}},\ \bibinfo
  {pages} {507} (\bibinfo {year} {2020})}\BibitemShut {NoStop}%
\bibitem [{\citenamefont {Andr{\'e}s}\ \emph {et~al.}(2020)\citenamefont
  {Andr{\'e}s}, \citenamefont {Su}, \citenamefont {Albert},\ and\ \citenamefont
  {Cator}}]{andresBuzzkillTargetingMosquito2020}%
  \BibitemOpen
  \bibfield  {author} {\bibinfo {author} {\bibfnamefont {M.}~\bibnamefont
  {Andr{\'e}s}}, \bibinfo {author} {\bibfnamefont {M.~P.}\ \bibnamefont {Su}},
  \bibinfo {author} {\bibfnamefont {J.}~\bibnamefont {Albert}},\ and\ \bibinfo
  {author} {\bibfnamefont {L.~J.}\ \bibnamefont {Cator}},\ }\bibfield  {title}
  {\bibinfo {title} {Buzzkill: Targeting the mosquito auditory system},\ }\href
  {https://doi.org/10.1016/j.cois.2020.04.003} {\bibfield  {journal} {\bibinfo
  {journal} {Curr. Opin. Insect Sci.}\ }\textbf {\bibinfo {volume} {40}},\
  \bibinfo {pages} {11} (\bibinfo {year} {2020})}\BibitemShut {NoStop}%
\bibitem [{\citenamefont {Boo}\ and\ \citenamefont
  {Richards}(1975)}]{booFineStructureScolopidia1975}%
  \BibitemOpen
  \bibfield  {author} {\bibinfo {author} {\bibfnamefont {K.~S.}\ \bibnamefont
  {Boo}}\ and\ \bibinfo {author} {\bibfnamefont {A.~G.}\ \bibnamefont
  {Richards}},\ }\bibfield  {title} {\bibinfo {title} {Fine structure of the
  scolopidia in the johnston's organ of male {{Aedes}} aegypti ({{L}}.)
  ({{Diptera}}: {{Culicidae}})},\ }\href
  {https://doi.org/10.1016/0020-7322(75)90031-8} {\bibfield  {journal}
  {\bibinfo  {journal} {Int. J. Insect Morphol. Embryol.}\ }\textbf {\bibinfo
  {volume} {4}},\ \bibinfo {pages} {549} (\bibinfo {year} {1975})}\BibitemShut
  {NoStop}%
\bibitem [{\citenamefont {Warren}\ \emph {et~al.}(2009)\citenamefont {Warren},
  \citenamefont {Gibson},\ and\ \citenamefont
  {Russell}}]{warrenSexRecognitionMidflight2009a}%
  \BibitemOpen
  \bibfield  {author} {\bibinfo {author} {\bibfnamefont {B.}~\bibnamefont
  {Warren}}, \bibinfo {author} {\bibfnamefont {G.}~\bibnamefont {Gibson}},\
  and\ \bibinfo {author} {\bibfnamefont {I.~J.}\ \bibnamefont {Russell}},\
  }\bibfield  {title} {\bibinfo {title} {Sex {{Recognition}} through
  {{Midflight Mating Duets}} in {{Culex Mosquitoes Is Mediated}} by {{Acoustic
  Distortion}}},\ }\href {https://doi.org/10.1016/j.cub.2009.01.059} {\bibfield
   {journal} {\bibinfo  {journal} {Curr. Biol.}\ }\textbf {\bibinfo {volume}
  {19}},\ \bibinfo {pages} {485} (\bibinfo {year} {2009})}\BibitemShut
  {NoStop}%
\bibitem [{\citenamefont {Sim{\~o}es}\ \emph {et~al.}(2016)\citenamefont
  {Sim{\~o}es}, \citenamefont {Ingham}, \citenamefont {Gibson},\ and\
  \citenamefont {Russell}}]{simoesRoleAcousticDistortion2016}%
  \BibitemOpen
  \bibfield  {author} {\bibinfo {author} {\bibfnamefont {P.~M.~V.}\
  \bibnamefont {Sim{\~o}es}}, \bibinfo {author} {\bibfnamefont {R.~A.}\
  \bibnamefont {Ingham}}, \bibinfo {author} {\bibfnamefont {G.}~\bibnamefont
  {Gibson}},\ and\ \bibinfo {author} {\bibfnamefont {I.~J.}\ \bibnamefont
  {Russell}},\ }\bibfield  {title} {\bibinfo {title} {A role for acoustic
  distortion in novel rapid frequency modulation behaviour in free-flying male
  mosquitoes},\ }\href {https://doi.org/10.1242/jeb.135293} {\bibfield
  {journal} {\bibinfo  {journal} {J. Exp. Biol.}\ }\textbf {\bibinfo {volume}
  {219}},\ \bibinfo {pages} {2039} (\bibinfo {year} {2016})}\BibitemShut
  {NoStop}%
\bibitem [{\citenamefont {Sim{\~o}es}\ \emph {et~al.}(2018)\citenamefont
  {Sim{\~o}es}, \citenamefont {Ingham}, \citenamefont {Gibson},\ and\
  \citenamefont {Russell}}]{simoesMaskingAuditoryBehaviour2018}%
  \BibitemOpen
  \bibfield  {author} {\bibinfo {author} {\bibfnamefont {P.~M.}\ \bibnamefont
  {Sim{\~o}es}}, \bibinfo {author} {\bibfnamefont {R.}~\bibnamefont {Ingham}},
  \bibinfo {author} {\bibfnamefont {G.}~\bibnamefont {Gibson}},\ and\ \bibinfo
  {author} {\bibfnamefont {I.~J.}\ \bibnamefont {Russell}},\ }\bibfield
  {title} {\bibinfo {title} {Masking of an auditory behaviour reveals how male
  mosquitoes use distortion to detect females},\ }\href
  {https://doi.org/10.1098/rspb.2017.1862} {\bibfield  {journal} {\bibinfo
  {journal} {Proc. Roy. Soc. B, Biol. Sci.}\ }\textbf {\bibinfo {volume}
  {285}},\ \bibinfo {pages} {11} (\bibinfo {year} {2018})}\BibitemShut
  {NoStop}%
\bibitem [{\citenamefont {Su}\ \emph {et~al.}(2018)\citenamefont {Su},
  \citenamefont {Andr{\'e}s}, \citenamefont {{Boyd-Gibbins}}, \citenamefont
  {Somers},\ and\ \citenamefont {Albert}}]{suSexSpeciesSpecific2018}%
  \BibitemOpen
  \bibfield  {author} {\bibinfo {author} {\bibfnamefont {M.~P.}\ \bibnamefont
  {Su}}, \bibinfo {author} {\bibfnamefont {M.}~\bibnamefont {Andr{\'e}s}},
  \bibinfo {author} {\bibfnamefont {N.}~\bibnamefont {{Boyd-Gibbins}}},
  \bibinfo {author} {\bibfnamefont {J.}~\bibnamefont {Somers}},\ and\ \bibinfo
  {author} {\bibfnamefont {J.~T.}\ \bibnamefont {Albert}},\ }\bibfield  {title}
  {\bibinfo {title} {Sex and species specific hearing mechanisms in mosquito
  flagellar ears},\ }\bibfield  {journal} {\bibinfo  {journal} {Nat. Commun.}\
  }\textbf {\bibinfo {volume} {9}},\ \href
  {https://doi.org/10.1038/s41467-018-06388-7} {10.1038/s41467-018-06388-7}
  (\bibinfo {year} {2018})\BibitemShut {NoStop}%
\bibitem [{\citenamefont {Faber}\ \emph
  {et~al.}(2025{\natexlab{a}})\citenamefont {Faber}, \citenamefont
  {Alampounti}, \citenamefont {Georgiades}, \citenamefont {Albert},\ and\
  \citenamefont {Bozovic}}]{faberMosquitoinspiredTheoreticalFramework2025a}%
  \BibitemOpen
  \bibfield  {author} {\bibinfo {author} {\bibfnamefont {J.}~\bibnamefont
  {Faber}}, \bibinfo {author} {\bibfnamefont {A.~C.}\ \bibnamefont
  {Alampounti}}, \bibinfo {author} {\bibfnamefont {M.}~\bibnamefont
  {Georgiades}}, \bibinfo {author} {\bibfnamefont {J.~T.}\ \bibnamefont
  {Albert}},\ and\ \bibinfo {author} {\bibfnamefont {D.}~\bibnamefont
  {Bozovic}},\ }\bibfield  {title} {\bibinfo {title} {A mosquito-inspired
  theoretical framework for acoustic signal detection},\ }\href
  {https://doi.org/10.1073/pnas.2500938122} {\bibfield  {journal} {\bibinfo
  {journal} {PNAS}\ }\textbf {\bibinfo {volume} {122}},\ \bibinfo {pages}
  {e2500938122} (\bibinfo {year} {2025}{\natexlab{a}})}\BibitemShut {NoStop}%
\bibitem [{\citenamefont {Schreiber}(2000)}]{Schreiber2000}%
  \BibitemOpen
  \bibfield  {author} {\bibinfo {author} {\bibfnamefont {T.}~\bibnamefont
  {Schreiber}},\ }\bibfield  {title} {\bibinfo {title} {Measuring information
  transfer},\ }\href {https://doi.org/10.1103/PhysRevLett.85.461} {\bibfield
  {journal} {\bibinfo  {journal} {Phys Rev Lett}\ }\textbf {\bibinfo {volume}
  {85}},\ \bibinfo {pages} {461} (\bibinfo {year} {2000})}\BibitemShut
  {NoStop}%
\bibitem [{\citenamefont {Arthur}\ \emph {et~al.}(2014)\citenamefont {Arthur},
  \citenamefont {Emr}, \citenamefont {Wyttenbach},\ and\ \citenamefont
  {Hoy}}]{arthurMosquitoAedesAegypti2014}%
  \BibitemOpen
  \bibfield  {author} {\bibinfo {author} {\bibfnamefont {B.~J.}\ \bibnamefont
  {Arthur}}, \bibinfo {author} {\bibfnamefont {K.~S.}\ \bibnamefont {Emr}},
  \bibinfo {author} {\bibfnamefont {R.~A.}\ \bibnamefont {Wyttenbach}},\ and\
  \bibinfo {author} {\bibfnamefont {R.~R.}\ \bibnamefont {Hoy}},\ }\bibfield
  {title} {\bibinfo {title} {Mosquito ( {{Aedes}} aegypti ) flight tones:
  {{Frequency}}, harmonicity, spherical spreading, and phase relationships},\
  }\href {https://doi.org/10.1121/1.4861233} {\bibfield  {journal} {\bibinfo
  {journal} {J. Acoust. Soc. Am.}\ }\textbf {\bibinfo {volume} {135}},\
  \bibinfo {pages} {933} (\bibinfo {year} {2014})}\BibitemShut {NoStop}%
\bibitem [{\citenamefont {Seo}\ \emph {et~al.}(2019)\citenamefont {Seo},
  \citenamefont {Hedrick},\ and\ \citenamefont
  {Mittal}}]{seoMechanismScalingWing2019}%
  \BibitemOpen
  \bibfield  {author} {\bibinfo {author} {\bibfnamefont {J.-H.}\ \bibnamefont
  {Seo}}, \bibinfo {author} {\bibfnamefont {T.~L.}\ \bibnamefont {Hedrick}},\
  and\ \bibinfo {author} {\bibfnamefont {R.}~\bibnamefont {Mittal}},\
  }\bibfield  {title} {\bibinfo {title} {Mechanism and scaling of wing tone
  generation in mosquitoes},\ }\href {https://doi.org/10.1088/1748-3190/ab54fc}
  {\bibfield  {journal} {\bibinfo  {journal} {Bioinspir. Biomim.}\ }\textbf
  {\bibinfo {volume} {15}},\ \bibinfo {pages} {016008} (\bibinfo {year}
  {2019})}\BibitemShut {NoStop}%
\bibitem [{\citenamefont
  {R{\"o}mer}(2020)}]{romerDirectionalHearingInsects2020}%
  \BibitemOpen
  \bibfield  {author} {\bibinfo {author} {\bibfnamefont {H.}~\bibnamefont
  {R{\"o}mer}},\ }\bibfield  {title} {\bibinfo {title} {Directional hearing in
  insects: Biophysical, physiological and ecological challenges},\ }\href
  {https://doi.org/10.1242/jeb.203224} {\bibfield  {journal} {\bibinfo
  {journal} {J. Exp. Biol.}\ }\textbf {\bibinfo {volume} {223}},\ \bibinfo
  {pages} {jeb203224} (\bibinfo {year} {2020})}\BibitemShut {NoStop}%
\bibitem [{\citenamefont {Faber}\ \emph
  {et~al.}(2025{\natexlab{b}})\citenamefont {Faber}, \citenamefont
  {Alampounti}, \citenamefont {Georgiades}, \citenamefont {Albert},\ and\
  \citenamefont {Bozovic}}]{faberAntennalBasedStrategiesSound2025}%
  \BibitemOpen
  \bibfield  {author} {\bibinfo {author} {\bibfnamefont {J.}~\bibnamefont
  {Faber}}, \bibinfo {author} {\bibfnamefont {A.~C.}\ \bibnamefont
  {Alampounti}}, \bibinfo {author} {\bibfnamefont {M.}~\bibnamefont
  {Georgiades}}, \bibinfo {author} {\bibfnamefont {J.~T.}\ \bibnamefont
  {Albert}},\ and\ \bibinfo {author} {\bibfnamefont {D.}~\bibnamefont
  {Bozovic}},\ }\href {https://doi.org/10.48550/arXiv.2505.04020} {\bibinfo
  {title} {Antennal-{{Based Strategies}} for {{Sound Localization}} by
  {{Insects}}}} (\bibinfo {year} {2025}{\natexlab{b}})\BibitemShut {NoStop}%
\bibitem [{\citenamefont
  {{Bennet-Clark}}(1971)}]{bennet-clarkAcousticsInsectSong1971}%
  \BibitemOpen
  \bibfield  {author} {\bibinfo {author} {\bibfnamefont {H.~C.}\ \bibnamefont
  {{Bennet-Clark}}},\ }\bibfield  {title} {\bibinfo {title} {Acoustics of
  {{Insect Song}}},\ }\href {https://doi.org/10.1038/234255a0} {\bibfield
  {journal} {\bibinfo  {journal} {Nature}\ }\textbf {\bibinfo {volume} {234}},\
  \bibinfo {pages} {255} (\bibinfo {year} {1971})}\BibitemShut {NoStop}%
\bibitem [{\citenamefont
  {{Bennet-Clark}}(1998)}]{bennet-clarkSizeScaleEffects1998}%
  \BibitemOpen
  \bibfield  {author} {\bibinfo {author} {\bibfnamefont {H.~C.}\ \bibnamefont
  {{Bennet-Clark}}},\ }\bibfield  {title} {\bibinfo {title} {Size and {{Scale
  Effects}} as {{Constraints}} in {{Insect Sound Communication}}},\ }\href@noop
  {} {\bibfield  {journal} {\bibinfo  {journal} {Philos. Trans. Biol. Sci.}\
  }\textbf {\bibinfo {volume} {353}},\ \bibinfo {pages} {407} (\bibinfo {year}
  {1998})}\BibitemShut {NoStop}%
\bibitem [{\citenamefont {Choe}\ \emph {et~al.}(1998)\citenamefont {Choe},
  \citenamefont {Magnasco},\ and\ \citenamefont
  {Hudspeth}}]{choeModelAmplificationHairbundle1998}%
  \BibitemOpen
  \bibfield  {author} {\bibinfo {author} {\bibfnamefont {Y.}~\bibnamefont
  {Choe}}, \bibinfo {author} {\bibfnamefont {M.~O.}\ \bibnamefont {Magnasco}},\
  and\ \bibinfo {author} {\bibfnamefont {A.~J.}\ \bibnamefont {Hudspeth}},\
  }\bibfield  {title} {\bibinfo {title} {A model for amplification of
  hair-bundle motion by cyclical binding of {{Ca2}}+ to
  mechanoelectrical-transduction channels},\ }\href
  {https://doi.org/10.1073/pnas.95.26.15321} {\bibfield  {journal} {\bibinfo
  {journal} {PNAS}\ }\textbf {\bibinfo {volume} {95}},\ \bibinfo {pages}
  {15321} (\bibinfo {year} {1998})}\BibitemShut {NoStop}%
\bibitem [{\citenamefont {Egu{\'i}luz}\ \emph {et~al.}(2000)\citenamefont
  {Egu{\'i}luz}, \citenamefont {Ospeck}, \citenamefont {Choe}, \citenamefont
  {Hudspeth},\ and\ \citenamefont
  {Magnasco}}]{eguiluzEssentialNonlinearitiesHearing2000}%
  \BibitemOpen
  \bibfield  {author} {\bibinfo {author} {\bibfnamefont {V.~M.}\ \bibnamefont
  {Egu{\'i}luz}}, \bibinfo {author} {\bibfnamefont {M.}~\bibnamefont {Ospeck}},
  \bibinfo {author} {\bibfnamefont {Y.}~\bibnamefont {Choe}}, \bibinfo {author}
  {\bibfnamefont {A.~J.}\ \bibnamefont {Hudspeth}},\ and\ \bibinfo {author}
  {\bibfnamefont {M.~O.}\ \bibnamefont {Magnasco}},\ }\bibfield  {title}
  {\bibinfo {title} {Essential {{Nonlinearities}} in {{Hearing}}},\ }\href
  {https://doi.org/10.1103/PhysRevLett.84.5232} {\bibfield  {journal} {\bibinfo
   {journal} {Phys. Rev. Lett.}\ }\textbf {\bibinfo {volume} {84}},\ \bibinfo
  {pages} {5232} (\bibinfo {year} {2000})}\BibitemShut {NoStop}%
\bibitem [{\citenamefont
  {Hudspeth}(2014)}]{hudspethIntegratingActiveProcess2014}%
  \BibitemOpen
  \bibfield  {author} {\bibinfo {author} {\bibfnamefont {A.~J.}\ \bibnamefont
  {Hudspeth}},\ }\bibfield  {title} {\bibinfo {title} {Integrating the active
  process of hair cells with cochlear function},\ }\href
  {https://doi.org/10.1038/nrn3786} {\bibfield  {journal} {\bibinfo  {journal}
  {Nat. Rev. Neurosci.}\ }\textbf {\bibinfo {volume} {15}},\ \bibinfo {pages}
  {600} (\bibinfo {year} {2014})}\BibitemShut {NoStop}%
\bibitem [{\citenamefont {Reichenbach}\ and\ \citenamefont
  {Hudspeth}(2014)}]{Reichenbach2014}%
  \BibitemOpen
  \bibfield  {author} {\bibinfo {author} {\bibfnamefont {T.}~\bibnamefont
  {Reichenbach}}\ and\ \bibinfo {author} {\bibfnamefont {a.~J.}\ \bibnamefont
  {Hudspeth}},\ }\bibfield  {title} {\bibinfo {title} {The physics of hearing:
  Fluid mechanics and the active process of the inner ear.},\ }\href
  {https://doi.org/10.1088/0034-4885/77/7/076601} {\bibfield  {journal}
  {\bibinfo  {journal} {Rep. Prog. Phys.}\ }\textbf {\bibinfo {volume} {77}},\
  \bibinfo {pages} {076601} (\bibinfo {year} {2014})}\BibitemShut {NoStop}%
\bibitem [{\citenamefont {Alonso}\ \emph {et~al.}(2025)\citenamefont {Alonso},
  \citenamefont {Gianoli}, \citenamefont {Fabella},\ and\ \citenamefont
  {Hudspeth}}]{alonsoAmplificationLocalCritical2025}%
  \BibitemOpen
  \bibfield  {author} {\bibinfo {author} {\bibfnamefont {R.~G.}\ \bibnamefont
  {Alonso}}, \bibinfo {author} {\bibfnamefont {F.}~\bibnamefont {Gianoli}},
  \bibinfo {author} {\bibfnamefont {B.}~\bibnamefont {Fabella}},\ and\ \bibinfo
  {author} {\bibfnamefont {A.~J.}\ \bibnamefont {Hudspeth}},\ }\bibfield
  {title} {\bibinfo {title} {Amplification through local critical behavior in
  the mammalian cochlea},\ }\href {https://doi.org/10.1073/pnas.2503389122}
  {\bibfield  {journal} {\bibinfo  {journal} {PNAS}\ }\textbf {\bibinfo
  {volume} {122}},\ \bibinfo {pages} {e2503389122} (\bibinfo {year}
  {2025})}\BibitemShut {NoStop}%
\bibitem [{\citenamefont {G{\"o}pfert}\ and\ \citenamefont
  {Robert}(2008)}]{gopfertActiveProcessesInsect2008}%
  \BibitemOpen
  \bibfield  {author} {\bibinfo {author} {\bibfnamefont {M.~C.}\ \bibnamefont
  {G{\"o}pfert}}\ and\ \bibinfo {author} {\bibfnamefont {D.}~\bibnamefont
  {Robert}},\ }\bibfield  {title} {\bibinfo {title} {Active {{Processes}} in
  {{Insect Hearing}}},\ }in\ \href
  {https://doi.org/10.1007/978-0-387-71469-1_6} {\emph {\bibinfo {booktitle}
  {Active {{Processes}} and {{Otoacoustic Emissions}} in {{Hearing}}}}},\
  \bibinfo {series and number} {Springer {{Handbook}} of {{Auditory
  Research}}},\ \bibinfo {editor} {edited by\ \bibinfo {editor} {\bibfnamefont
  {G.~A.}\ \bibnamefont {Manley}}, \bibinfo {editor} {\bibfnamefont {R.~R.}\
  \bibnamefont {Fay}},\ and\ \bibinfo {editor} {\bibfnamefont {A.~N.}\
  \bibnamefont {Popper}}}\ (\bibinfo  {publisher} {Springer},\ \bibinfo
  {address} {New York, NY},\ \bibinfo {year} {2008})\ pp.\ \bibinfo {pages}
  {191--209}\BibitemShut {NoStop}%
\bibitem [{\citenamefont {G{\"o}pfert}\ and\ \citenamefont
  {Hennig}(2016)}]{gopfertHearingInsects2016}%
  \BibitemOpen
  \bibfield  {author} {\bibinfo {author} {\bibfnamefont {M.~C.}\ \bibnamefont
  {G{\"o}pfert}}\ and\ \bibinfo {author} {\bibfnamefont {R.~M.}\ \bibnamefont
  {Hennig}},\ }\bibfield  {title} {\bibinfo {title} {Hearing in {{Insects}}},\
  }\href {https://doi.org/10.1146/annurev-ento-010715-023631} {\bibfield
  {journal} {\bibinfo  {journal} {Annu. Rev. Entomol.}\ }\textbf {\bibinfo
  {volume} {61}},\ \bibinfo {pages} {257} (\bibinfo {year} {2016})}\BibitemShut
  {NoStop}%
\bibitem [{\citenamefont {Nadrowski}\ \emph {et~al.}(2011)\citenamefont
  {Nadrowski}, \citenamefont {Effertz}, \citenamefont {Senthilan},\ and\
  \citenamefont {G{\"o}pfert}}]{nadrowskiAntennalHearingInsects2011}%
  \BibitemOpen
  \bibfield  {author} {\bibinfo {author} {\bibfnamefont {B.}~\bibnamefont
  {Nadrowski}}, \bibinfo {author} {\bibfnamefont {T.}~\bibnamefont {Effertz}},
  \bibinfo {author} {\bibfnamefont {P.~R.}\ \bibnamefont {Senthilan}},\ and\
  \bibinfo {author} {\bibfnamefont {M.~C.}\ \bibnamefont {G{\"o}pfert}},\
  }\bibfield  {title} {\bibinfo {title} {Antennal hearing in insects - {{New}}
  findings, new questions},\ }\href
  {https://doi.org/10.1016/j.heares.2010.03.092} {\bibfield  {journal}
  {\bibinfo  {journal} {Hear. Res.}\ }\textbf {\bibinfo {volume} {273}},\
  \bibinfo {pages} {7} (\bibinfo {year} {2011})}\BibitemShut {NoStop}%
\bibitem [{\citenamefont {Mhatre}(2015)}]{mhatreActiveAmplificationInsect2015}%
  \BibitemOpen
  \bibfield  {author} {\bibinfo {author} {\bibfnamefont {N.}~\bibnamefont
  {Mhatre}},\ }\bibfield  {title} {\bibinfo {title} {Active amplification in
  insect ears: Mechanics, models and molecules},\ }\href
  {https://doi.org/10.1007/s00359-014-0969-0} {\bibfield  {journal} {\bibinfo
  {journal} {J. Comp. Physiol. A}\ }\textbf {\bibinfo {volume} {201}},\
  \bibinfo {pages} {19} (\bibinfo {year} {2015})}\BibitemShut {NoStop}%
\bibitem [{\citenamefont {Albert}\ and\ \citenamefont
  {Kozlov}(2016)}]{albertComparativeAspectsHearing2016}%
  \BibitemOpen
  \bibfield  {author} {\bibinfo {author} {\bibfnamefont {J.~T.}\ \bibnamefont
  {Albert}}\ and\ \bibinfo {author} {\bibfnamefont {A.~S.}\ \bibnamefont
  {Kozlov}},\ }\bibfield  {title} {\bibinfo {title} {Comparative {{Aspects}} of
  {{Hearing}} in {{Vertebrates}} and {{Insects}} with {{Antennal Ears}}},\
  }\href {https://doi.org/10.1016/j.cub.2016.09.017} {\bibfield  {journal}
  {\bibinfo  {journal} {Curr. Biol.}\ }\textbf {\bibinfo {volume} {26}},\
  \bibinfo {pages} {R1050} (\bibinfo {year} {2016})}\BibitemShut {NoStop}%
\bibitem [{\citenamefont {Jakhete}\ \emph {et~al.}(2017)\citenamefont
  {Jakhete}, \citenamefont {Allan},\ and\ \citenamefont
  {Mankin}}]{jakheteWingbeatFrequencySweepVisual2017}%
  \BibitemOpen
  \bibfield  {author} {\bibinfo {author} {\bibfnamefont {S.~S.}\ \bibnamefont
  {Jakhete}}, \bibinfo {author} {\bibfnamefont {S.~A.}\ \bibnamefont {Allan}},\
  and\ \bibinfo {author} {\bibfnamefont {R.~W.}\ \bibnamefont {Mankin}},\
  }\bibfield  {title} {\bibinfo {title} {Wingbeat {{Frequency-Sweep}} and
  {{Visual Stimuli}} for {{Trapping Male Aedes}} aegypti ({{Diptera}}:
  {{Culicidae}})},\ }\href {https://doi.org/10.1093/jme/tjx074} {\bibfield
  {journal} {\bibinfo  {journal} {J. Med. Entomol.}\ }\textbf {\bibinfo
  {volume} {54}},\ \bibinfo {pages} {1415} (\bibinfo {year}
  {2017})}\BibitemShut {NoStop}%
\bibitem [{\citenamefont {Ogawa}(1988)}]{ogawaFieldStudyAcoustic1988}%
  \BibitemOpen
  \bibfield  {author} {\bibinfo {author} {\bibfnamefont {K.-i.}\ \bibnamefont
  {Ogawa}},\ }\bibfield  {title} {\bibinfo {title} {Field {{Study}} on
  {{Acoustic Trapping}} of {{Mansonia}} ({{Diptera}} : {{Culicidae}}) in
  {{Malaysia I}}. {{Mass-Trapping}} of {{Males}} by a {{Cylindrical Sound
  Trap}}},\ }\href {https://doi.org/10.1303/aez.23.265} {\bibfield  {journal}
  {\bibinfo  {journal} {Appl. Entomol. Zool.}\ }\textbf {\bibinfo {volume}
  {23}},\ \bibinfo {pages} {265} (\bibinfo {year} {1988})}\BibitemShut
  {NoStop}%
\bibitem [{\citenamefont {Ikeshoji}\ and\ \citenamefont
  {Yap}(1990)}]{ikeshojiImpactInsecticidetreatedSound1990}%
  \BibitemOpen
  \bibfield  {author} {\bibinfo {author} {\bibfnamefont {T.}~\bibnamefont
  {Ikeshoji}}\ and\ \bibinfo {author} {\bibfnamefont {H.}~\bibnamefont {Yap}},\
  }\bibfield  {title} {\bibinfo {title} {Impact of the insecticide-treated
  sound traps on an {{Aedes}} albopictus population},\ }\href
  {https://doi.org/10.7601/mez.41.213} {\bibfield  {journal} {\bibinfo
  {journal} {Med. Entomol. Zool.}\ }\textbf {\bibinfo {volume} {41}},\ \bibinfo
  {pages} {213} (\bibinfo {year} {1990})}\BibitemShut {NoStop}%
\bibitem [{\citenamefont {Stone}\ \emph {et~al.}(2013)\citenamefont {Stone},
  \citenamefont {Tuten},\ and\ \citenamefont
  {Dobson}}]{stoneDeterminantsMaleAedes2013}%
  \BibitemOpen
  \bibfield  {author} {\bibinfo {author} {\bibfnamefont {C.~M.}\ \bibnamefont
  {Stone}}, \bibinfo {author} {\bibfnamefont {H.~C.}\ \bibnamefont {Tuten}},\
  and\ \bibinfo {author} {\bibfnamefont {S.~L.}\ \bibnamefont {Dobson}},\
  }\bibfield  {title} {\bibinfo {title} {Determinants of {{Male Aedes}} aegypti
  and {{Aedes}} polynesiensis ({{Diptera}}: {{Culicidae}}) {{Response}} to
  {{Sound}}: {{Efficacy}} and {{Considerations}} for {{Use}} of {{Sound Traps}}
  in the {{Field}}},\ }\href {https://doi.org/10.1603/ME13023} {\bibfield
  {journal} {\bibinfo  {journal} {J. Med. Entomol.}\ }\textbf {\bibinfo
  {volume} {50}},\ \bibinfo {pages} {723} (\bibinfo {year} {2013})}\BibitemShut
  {NoStop}%
\bibitem [{\citenamefont {Johnson}\ and\ \citenamefont
  {Ritchie}(2016)}]{johnsonSirensSongExploitation2016}%
  \BibitemOpen
  \bibfield  {author} {\bibinfo {author} {\bibfnamefont {B.~J.}\ \bibnamefont
  {Johnson}}\ and\ \bibinfo {author} {\bibfnamefont {S.~A.}\ \bibnamefont
  {Ritchie}},\ }\bibfield  {title} {\bibinfo {title} {The {{Siren}}'s {{Song}}:
  {{Exploitation}} of {{Female Flight Tones}} to {{Passively Capture Male
  Aedes}} aegypti ({{Diptera}}: {{Culicidae}})},\ }\href
  {https://doi.org/10.1093/jme/tjv165} {\bibfield  {journal} {\bibinfo
  {journal} {J. Med. Entomol.}\ }\textbf {\bibinfo {volume} {53}},\ \bibinfo
  {pages} {245} (\bibinfo {year} {2016})}\BibitemShut {NoStop}%
\bibitem [{\citenamefont {Rohde}\ \emph {et~al.}(2019)\citenamefont {Rohde},
  \citenamefont {Staunton}, \citenamefont {Zeak}, \citenamefont {Beebe},
  \citenamefont {Snoad}, \citenamefont {Bondarenco}, \citenamefont
  {Liddington}, \citenamefont {Anderson}, \citenamefont {Xiang}, \citenamefont
  {Mankin},\ and\ \citenamefont
  {Ritchie}}]{rohdeWaterproofLowcostLongbatterylife2019}%
  \BibitemOpen
  \bibfield  {author} {\bibinfo {author} {\bibfnamefont {B.~B.}\ \bibnamefont
  {Rohde}}, \bibinfo {author} {\bibfnamefont {K.~M.}\ \bibnamefont {Staunton}},
  \bibinfo {author} {\bibfnamefont {N.~C.}\ \bibnamefont {Zeak}}, \bibinfo
  {author} {\bibfnamefont {N.}~\bibnamefont {Beebe}}, \bibinfo {author}
  {\bibfnamefont {N.}~\bibnamefont {Snoad}}, \bibinfo {author} {\bibfnamefont
  {A.}~\bibnamefont {Bondarenco}}, \bibinfo {author} {\bibfnamefont
  {C.}~\bibnamefont {Liddington}}, \bibinfo {author} {\bibfnamefont {J.~A.}\
  \bibnamefont {Anderson}}, \bibinfo {author} {\bibfnamefont {W.}~\bibnamefont
  {Xiang}}, \bibinfo {author} {\bibfnamefont {R.~W.}\ \bibnamefont {Mankin}},\
  and\ \bibinfo {author} {\bibfnamefont {S.~A.}\ \bibnamefont {Ritchie}},\
  }\bibfield  {title} {\bibinfo {title} {Waterproof, low-cost,
  long-battery-life sound trap for surveillance of male {{Aedes}} aegypti for
  rear-and-release mosquito control programmes},\ }\href
  {https://doi.org/10.1186/s13071-019-3647-9} {\bibfield  {journal} {\bibinfo
  {journal} {Parasit. Vectors}\ }\textbf {\bibinfo {volume} {12}},\ \bibinfo
  {pages} {417} (\bibinfo {year} {2019})}\BibitemShut {NoStop}%
\bibitem [{\citenamefont {Dieng}\ \emph {et~al.}(2019)\citenamefont {Dieng},
  \citenamefont {The}, \citenamefont {Satho}, \citenamefont {Miake},
  \citenamefont {Wydiamala}, \citenamefont {Kassim}, \citenamefont {Hashim},
  \citenamefont {Morales~Vargas},\ and\ \citenamefont
  {Morales}}]{diengElectronicSongScary2019}%
  \BibitemOpen
  \bibfield  {author} {\bibinfo {author} {\bibfnamefont {H.}~\bibnamefont
  {Dieng}}, \bibinfo {author} {\bibfnamefont {C.~C.}\ \bibnamefont {The}},
  \bibinfo {author} {\bibfnamefont {T.}~\bibnamefont {Satho}}, \bibinfo
  {author} {\bibfnamefont {F.}~\bibnamefont {Miake}}, \bibinfo {author}
  {\bibfnamefont {E.}~\bibnamefont {Wydiamala}}, \bibinfo {author}
  {\bibfnamefont {N.~F.~A.}\ \bibnamefont {Kassim}}, \bibinfo {author}
  {\bibfnamefont {N.~A.}\ \bibnamefont {Hashim}}, \bibinfo {author}
  {\bibfnamefont {R.~E.}\ \bibnamefont {Morales~Vargas}},\ and\ \bibinfo
  {author} {\bibfnamefont {N.~P.}\ \bibnamefont {Morales}},\ }\bibfield
  {title} {\bibinfo {title} {The electronic song ``{{Scary Monsters}} and
  {{Nice Sprites}}'' reduces host attack and mating success in the dengue
  vector {{{\emph{Aedes}}}}{\emph{ aegypti}}},\ }\href
  {https://doi.org/10.1016/j.actatropica.2019.03.027} {\bibfield  {journal}
  {\bibinfo  {journal} {Acta Tropica}\ }\textbf {\bibinfo {volume} {194}},\
  \bibinfo {pages} {93} (\bibinfo {year} {2019})}\BibitemShut {NoStop}%
\bibitem [{\citenamefont {Ni}\ \emph {et~al.}(2025)\citenamefont {Ni},
  \citenamefont {Kassim}, \citenamefont {Ayub}, \citenamefont {Abuelmaali},
  \citenamefont {Mashlawi},\ and\ \citenamefont
  {Dieng}}]{niImpactDiverseMusical2025}%
  \BibitemOpen
  \bibfield  {author} {\bibinfo {author} {\bibfnamefont {C.~Y.}\ \bibnamefont
  {Ni}}, \bibinfo {author} {\bibfnamefont {N.~F.~A.}\ \bibnamefont {Kassim}},
  \bibinfo {author} {\bibfnamefont {N.~M.}\ \bibnamefont {Ayub}}, \bibinfo
  {author} {\bibfnamefont {S.~A.}\ \bibnamefont {Abuelmaali}}, \bibinfo
  {author} {\bibfnamefont {A.~M.}\ \bibnamefont {Mashlawi}},\ and\ \bibinfo
  {author} {\bibfnamefont {H.}~\bibnamefont {Dieng}},\ }\bibfield  {title}
  {\bibinfo {title} {Impact of diverse musical genres on blood-feeding and
  mating behavior in {{Aedes}} aegypti mosquitoes},\ }\href
  {https://doi.org/10.4103/JVBD.JVBD_111_24} {\bibfield  {journal} {\bibinfo
  {journal} {J Vector Borne Dis}\ }\textbf {\bibinfo {volume} {62}},\ \bibinfo
  {pages} {211} (\bibinfo {year} {2025})}\BibitemShut {NoStop}%
\bibitem [{\citenamefont {Skolnik}(1970)}]{SkolnikRadar1970}%
  \BibitemOpen
  \bibfield  {author} {\bibinfo {author} {\bibfnamefont {M.~I.}\ \bibnamefont
  {Skolnik}},\ }\href@noop {} {\emph {\bibinfo {title} {Radar Handbook}}}\
  (\bibinfo  {publisher} {McGraw Hill},\ \bibinfo {year} {1970})\BibitemShut
  {NoStop}%
\bibitem [{\citenamefont {Somers}\ \emph {et~al.}(2022)\citenamefont {Somers},
  \citenamefont {Georgiades}, \citenamefont {Su}, \citenamefont {Bagi},
  \citenamefont {Andr{\'e}s}, \citenamefont {Alampounti}, \citenamefont
  {Mills}, \citenamefont {Ntabaliba}, \citenamefont {Moore}, \citenamefont
  {Spaccapelo},\ and\ \citenamefont {Albert}}]{somersHittingRightNote2022}%
  \BibitemOpen
  \bibfield  {author} {\bibinfo {author} {\bibfnamefont {J.}~\bibnamefont
  {Somers}}, \bibinfo {author} {\bibfnamefont {M.}~\bibnamefont {Georgiades}},
  \bibinfo {author} {\bibfnamefont {M.~P.}\ \bibnamefont {Su}}, \bibinfo
  {author} {\bibfnamefont {J.}~\bibnamefont {Bagi}}, \bibinfo {author}
  {\bibfnamefont {M.}~\bibnamefont {Andr{\'e}s}}, \bibinfo {author}
  {\bibfnamefont {A.}~\bibnamefont {Alampounti}}, \bibinfo {author}
  {\bibfnamefont {G.}~\bibnamefont {Mills}}, \bibinfo {author} {\bibfnamefont
  {W.}~\bibnamefont {Ntabaliba}}, \bibinfo {author} {\bibfnamefont {S.~J.}\
  \bibnamefont {Moore}}, \bibinfo {author} {\bibfnamefont {R.}~\bibnamefont
  {Spaccapelo}},\ and\ \bibinfo {author} {\bibfnamefont {J.~T.}\ \bibnamefont
  {Albert}},\ }\bibfield  {title} {\bibinfo {title} {Hitting the right note at
  the right time: {{Circadian}} control of audibility in {{Anopheles}} mosquito
  mating swarms is mediated by flight tones},\ }\href
  {https://doi.org/10.1126/sciadv.abl4844} {\bibfield  {journal} {\bibinfo
  {journal} {Sci. Adv.}\ }\textbf {\bibinfo {volume} {8}},\ \bibinfo {pages}
  {eabl4844} (\bibinfo {year} {2022})}\BibitemShut {NoStop}%
\bibitem [{\citenamefont {Faber}\ \emph
  {et~al.}(2025{\natexlab{c}})\citenamefont {Faber}, \citenamefont
  {Alampounti}, \citenamefont {Georgiades}, \citenamefont {Albert},\ and\
  \citenamefont {Bozovic}}]{DataAvailability}%
  \BibitemOpen
  \bibfield  {author} {\bibinfo {author} {\bibfnamefont {J.}~\bibnamefont
  {Faber}}, \bibinfo {author} {\bibfnamefont {A.~C.}\ \bibnamefont
  {Alampounti}}, \bibinfo {author} {\bibfnamefont {M.}~\bibnamefont
  {Georgiades}}, \bibinfo {author} {\bibfnamefont {J.~T.}\ \bibnamefont
  {Albert}},\ and\ \bibinfo {author} {\bibfnamefont {D.}~\bibnamefont
  {Bozovic}},\ }\href@noop {} {\bibinfo {title} {Python code for all analysis
  and figure generation}},\ \bibinfo {howpublished}
  {\href{https://github.com/jfaber3/Sound-Masking-Strategies}{Available on
  GitHub}} (\bibinfo {year} {2025}{\natexlab{c}})\BibitemShut {NoStop}%
\bibitem [{\citenamefont {Bossomaier}\ \emph {et~al.}(2016)\citenamefont
  {Bossomaier}, \citenamefont {Barnett}, \citenamefont {Harr\'e},\ and\
  \citenamefont {Lizier}}]{BossomaierTransferEntropy2016}%
  \BibitemOpen
  \bibfield  {author} {\bibinfo {author} {\bibfnamefont {T.}~\bibnamefont
  {Bossomaier}}, \bibinfo {author} {\bibfnamefont {L.}~\bibnamefont {Barnett}},
  \bibinfo {author} {\bibfnamefont {M.}~\bibnamefont {Harr\'e}},\ and\ \bibinfo
  {author} {\bibfnamefont {J.~T.}\ \bibnamefont {Lizier}},\ }\href
  {https://doi.org/10.1007/978-3-319-43222-9} {\emph {\bibinfo {title} {An
  Introduction to Transfer Entropy}}}\ (\bibinfo  {publisher} {Springer Cham},\
  \bibinfo {year} {2016})\BibitemShut {NoStop}%
\end{thebibliography}%

\section*{Appendix A: Transfer Entropy}

The transfer entropy \cite{Schreiber2000, BossomaierTransferEntropy2016} from process $J$ to process $I$ is defined as

\begin{eqnarray}
T_{J \rightarrow I} = \sum p(i_{n+1}, i_{n}^{(k)}, j_{n}^{(l)})
\log \frac {p(i_{n+1} \ | \ i_{n}^{(k)}, j_{n}^{(l)})} {p(i_{n+1} \ | \ i_{n}^{(k)})}, \hspace{1cm}
\label{eq:TE}
\end{eqnarray}

\noindent where $i_{n}^{(k)} = (i_n,...,i_{n-k+1})$ are the $k$ most recent states of process $I$. Therefore, $p(i_{n+1} \ | \ i_{n}^{(k)}, j_{n}^{(l)})$ is the conditional probability of finding process $I$ in state $i_{n+1}$ at time $n+1$, given that the previous $k$ states of process $I$ were $i_{n}^{(k)}$ and that the previous $l$ states of process $J$ were $j_{n}^{(l)}$. The summation runs over all accessible states for both processes and over all points in the time series. The transfer entropy measures how much one's ability to predict the future of process $I$ is improved upon learning the history of process $J$. It has been used extensively as an information theoretic measure of signal detection \cite{BossomaierTransferEntropy2016}.

For our application, we seek to measure the transfer entropy from the target signal, $F_f(t)$, to the response of the system ($z(t)$ for model 1 or $z_2(t)$ for model 2). We compute the transfer entropy once using just the real parts of the signals, and again using just the imaginary parts. As these values were nearly identical in all cases, we simply report the mean of the two. The calculations were carried out by first discretizing the signals into one of two states at every point in time. We use the mean value of each signal as the delineation between the two states. The choice of $k$ and $l$ has little effect on the results, so we select $k = l = 5$, and sample the 5 points such that they span one mean period of $F_f(t)$. A transfer entropy of zero indicates that the two processes are completely independent and that no signal detection has occurred. Since we chose two states in the discretization, the maximum value this measure can take is $1$ bit.

\clearpage

\section*{Appendix B: Additional frequency-modulation (FM) Masks}

\begin{figure}[h!]
\includegraphics[width=\columnwidth]{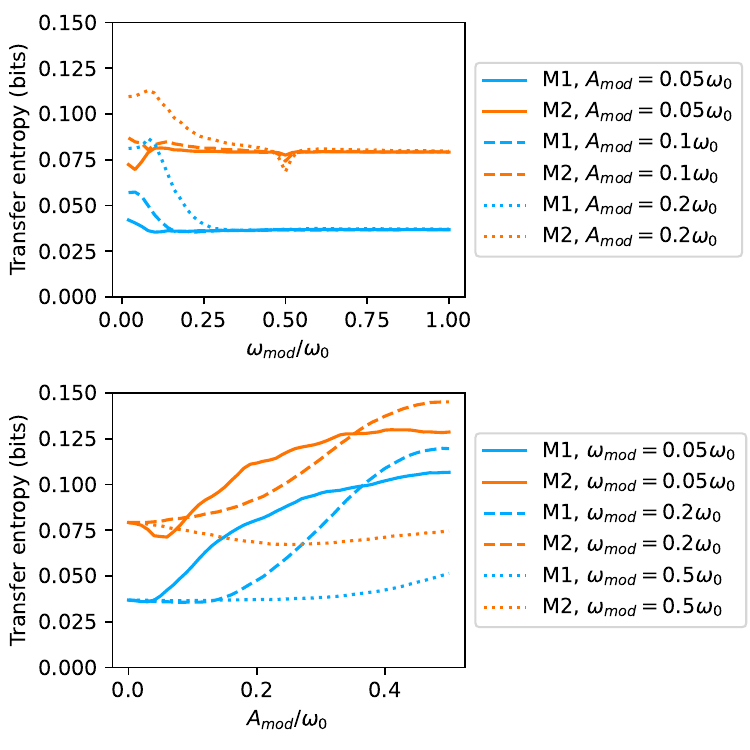}
\caption{\textbf{Sinusoidal FM mask.} The effects of the masking parameters, $A_{mod}$ and $\omega_{mod}$, on the transfer entropy from $F_f(t)$ to $z(t)$ (model 1: M1), and from $F_f(t)$ to $z_2(t)$ (model 2: M2).}
\label{figS1}
\end{figure}

\begin{figure}[h!]
\includegraphics[width=\columnwidth]{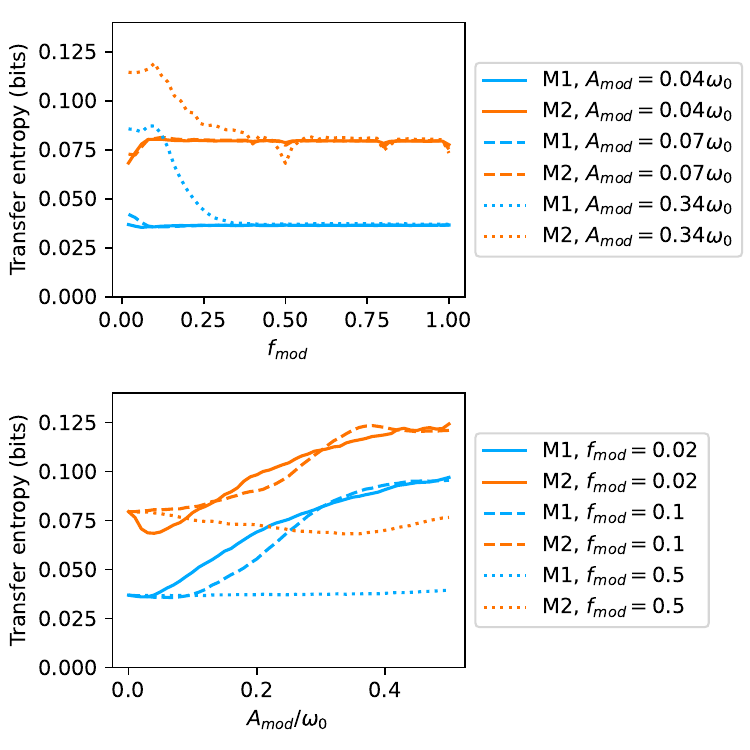}
\caption{\textbf{Sawtooth FM mask.} The effects of the masking parameters, $f_{mod}$ and $A_{mod}$, on the transfer entropy from $F_f(t)$ to $z(t)$ (model 1: M1), and from $F_f(t)$ to $z_2(t)$ (model 2: M2).}
\label{figS2}
\end{figure}

\begin{figure}[h!]
\includegraphics[width=\columnwidth]{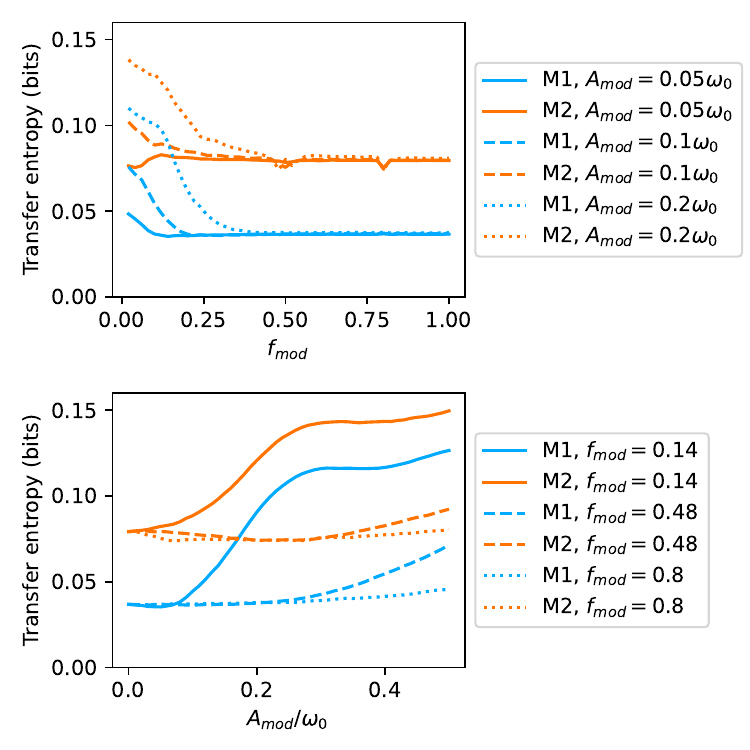}
\caption{\textbf{Square-wave FM mask.} The effects of the masking parameters, $f_{mod}$ and $A_{mod}$, on the transfer entropy from $F_f(t)$ to $z(t)$ (model 1: M1), and from $F_f(t)$ to $z_2(t)$ (model 2: M2).}
\label{figS3}
\end{figure}

We now test the effectiveness of FM masks with alternative modulators. In FIG. \ref{figS1}, we plot the transfer entropy for sinusoidal modulators of the form, $m(t) = A_{mod}\sin(\omega_{mod}t)$. We also show the results for sawtooth modulators (FIG. \ref{figS2}) and square-wave modulators (FIG. \ref{figS3}). Note that these are limiting cases of Eq. \ref{eq:fm4}, where $\alpha = 1$ produces sawtooth modulations and $\alpha = 0$ produces square-wave modulations. All three cases qualitatively show the same behavior, where effective masks can be found when using rapid frequency modulations (high $f_{mod}$) with small $A_{mod}$.

\newpage

\newpage

\end{document}